\documentclass[prb,twocolumn,showpacs,amsmath,amssymb]{revtex4-2}
\usepackage{graphicx}
\usepackage{color}
\usepackage{xcolor}
\usepackage{bm,upgreek}

\usepackage{fancyhdr}

\definecolor{grey}{rgb}{.65,.65,.65}

\begin{document}
\title{Two-fluid model analysis of the terahertz conductivity of YBaCuO samples: optimally doped, underdoped and overdoped cases}

\author{Michal~\v{S}indler$^1$}
\author{Wen-Yen~Tzeng$^2$}
\author{Chih-Wei~Luo$^{3-5}$}
\author{Jiunn-Yuan Lin$^5$}
\author{Christelle~Kadlec$^1$}

\affiliation{$^1$Institute of Physics, Academy of Sciences of the Czech Republic, Na Slovance 1999/2, 18200 Prague 8, Czech Republic}
\affiliation{$^2$Department of Electronic Engineering, National Formosa University, Yunlin 632, Taiwan}
\affiliation{$^3$Department of Electrophysics, National Yang Ming Chiao Tung University, Hsinchu 30010, Taiwan}
\affiliation{$^4$National Synchrotron Radiation Research Center, Hsinchu, 30076, Taiwan}
\affiliation{$^5$Institute of Physics, National Yang Ming Chiao Tung University, Hsinchu 30010, Taiwan}
\date{\today}




\begin{abstract}
The complex conductivity of underdoped and optimally doped YBa$_2$Cu$_3$O$_{7-\delta}$ samples and overdoped similar compound Y$_{0.7}$Ca$_{0.3}$Ba$_2$Cu$_3$O$_{7-\delta}$ was
 measured using time-domain terahertz spectroscopy. In the normal state, the frequency dependence is described by the Drude model. Below the critical temperature $T_\mathrm{c}$, the two-fluid model was successfully employed to fit all the spectra, from 5 K up to $T_\mathrm{c}$. The temperature behaviour of fundamental parameters such as the scattering rate $1/\tau$, the superfluid (normal) fraction $f_\mathrm{s}$ ($f_\mathrm{n}$) and the conductivity $\sigma$ was investigated at given frequencies. 
 For the optimally doped and the overdoped
samples, even at 5 K, a fifth of the electrons do not condense to the superfluid fraction.
  We observed that a substantial fraction of electrons do not condense to the superfluid fraction even at 5 K for optimally doped and overdoped samples.
The real part of the conductivity $\sigma_1(T)$ exhibits a peak at low frequencies. It can be observed for all three stoichiometries and its exact shape depends on the quality of the sample. A further analysis shows that this peak is a consequence of the competition between the scattering time $\tau(T)$ and the superfluid fraction $f_\mathrm{s}(T)$.
\end{abstract}

\maketitle

\section{Introduction}

After nearly four decades of dedicated research, the scientific community continues to grapple with a comprehensive understanding of the cuprate material family. Among this family, one of the most extensively studied high-temperature superconductors is YBa$_2$Cu$_3$O$_{7-\delta}$ (YBCO). The phase diagram of YBCO~\cite{Taillefer_review} is intricately linked to the precise stoichiometry of oxygen atoms within its structure, as well as the distribution of these atoms at a microscopic level within the unit cell. Even with identical oxygen stoichiometry, variations in properties can arise due to differences in oxygen distribution between planes and chains~\cite{Liang2006}.

A critical parameter that governs the phase diagram of YBCO is the hole doping level, represented as $p$, which denotes the number of holes per one cuprate plane. As the hole doping level increases, the material transitions from an antiferromagnetic phase to a pseudogap phase, and eventually to a strange metal phase. Below 100 K at ambient pressure, a superconducting dome is observed, beginning at $p$ = 0.05 and reaching its maximum critical temperature $T_c$ at $p$ = 0.16. However, at higher levels of hole doping, YBCO compounds become progressively chemically unstable. To investigate the phase diagram in the overdoped regime, researchers have explored altering the chemical composition by partially replacing Y with a lighter element such as Ca. This modification allows for the study of YBCO compounds beyond the limits of their chemical stability.

In order to deepen our understanding of the superconducting state in high-temperature superconductors, it is imperative to investigate low-energy excitations below the optical gap. Previous studies have extensively examined YBCO, including both crystal samples and thin films, using microwave~\cite{Bonn1993,Bonn1994,Gao1993,Vaulchier1996}  and terahertz~\cite{Nuss1991,Ludwig1996,Frenkel1996,Pimenov1999,Pimenov2000,Tsai2003} methods. Despite the wealth of experimental data available, there remains a scarcity of research on the electrodynamic properties of YBCO at varying levels of hole doping~\cite{Ludwig1996}, particularly for the overdoped case.

In our work we prepared several YBCO samples with different doping ranging from underdoped to optimally doped and finally overdoped case in the essentially same compound Y$_{0.7}$Ca$_{0.3}$Ba$_2$Cu$_3$O$_{7-\delta}$ (YCBCO).  The Ca doping introduces some out of plane defect.
  From our terahertz measurements, we obtain temperature dependence of the complex conductivity, the scattering rate and the density of charge carriers for different levels of hole doping. 
These results offer insights into the nature of pairing in the superconducting state and shed light on whether quasiparticles persist up to the zero-temperature limit.

\section {Theoretical description}
As currently, there is no universally accepted microscopic theoretical description of high-temperature superconductors, we use the phenomenological description by the two-fluid model~\cite{Gorter-Casimir} in the form given by Pimenov~\cite{Pimenov1999}

\begin{equation}
\sigma(\omega)=   \frac{n_\mathrm{s}e^2}{m} \left( \pi \delta(\omega) +\mathrm{i} \frac{1}{\omega}  \right)+   \frac{n_\mathrm{n}e^2}{m} \frac{\tau}{1-i\omega\tau},
\label{2f_model}
\end{equation}
where  the superfluid concentration $n_\mathrm{s}$  and   the normal fluid concentration $n_\mathrm{n}$ are temperature dependent, but the total electron concentration $n=n_\mathrm{s}+n_\mathrm{n}$  is constant,  $e$ is the elementary charge, $1/\tau$ is the scattering rate, $m$ is the effective electron mass, $\delta(\omega)$ is $\delta$-function and $\omega$ is the circular frequency related to the frequency $\nu$ by the relation $\omega=2\pi \nu$. While it is customary to refer to frequency $\nu$ in the experiment, in the theory $\omega$ is used. 
The first term is the London description of the complex conductivity and the second term describes the normal fluid by using the Drude model.  Above the critical temperature $T_c$,  $n_\mathrm{s}$ is zero and $n_\mathrm{n}=n$.
In the  DC limit, the normal-state resistivity is $\varrho=1/\sigma=m/(ne^2\tau)$ thus the temperature dependence of the DC resistivity is driven by the temperature behaviour of the scattering rate.

The superfluid response allows us to determine the London penetration depth $\lambda_L$:

\begin{equation}
\lambda_L^2= \frac{m}{\mu_0 n_\mathrm{s} e^2}
\label{london_eq}
\end{equation}
where $\mu_0$ is the permeability of vacuum. The penetration depth
is related to the pairing mechanism. For classical superconductors, its temperature dependence is described by the BCS theory~\cite{BCS}.  In the case of the d-wave pairing which is more appropriate for cuprates, theories predict a linear temperature dependence of the penetration depth, at least in the $T\rightarrow 0$ limit~\cite{Annett1991}. A quadratic dependence can be obtained for d-wave superconductors with strong impurity scattering~\cite{Prohammer1991}. Phenomenologically, a temperature dependence $1-(T/T_\mathrm{c})^4$ was suggested by Gorter and Casimir~\cite{Gorter-Casimir}. In order to distinguish relevant cases, we use the temperature dependence of the penetration depth by the formula~\cite{Brorson1996}
\begin{equation}
\left [ \frac{\lambda_L(0)}{\lambda_L(T)} \right]^2 = 1-  t^{\alpha} ,
\label{lambda_T_dep}
\end{equation}
where  $\lambda_L(0)$ is the London penetration depth at the zero temperature limit, $t$ is the reduced temperature $T/T_c$ and the formula includes all mentioned cases for exponent $\alpha$ equal to 1, 2 and 4.

\section{Experiment}

\subsection{Sample preparation}

The YBa$_2$Cu$_3$O$_{7-\delta}$ (YBCO) and Y$_{0.7}$Ca$_{0.3}$Ba$_2$ Cu$_3$O$_{7-\delta}$ (YCBCO) samples used in this study were prepared by pulsed laser deposition (PLD).
 PLD has an overwhelming superiority over other deposition methods in high-Tc superconducting thin films, since it can rapidly produce high-quality samples. Indeed, YBCO thin films previously used to study optical spectroscopy were often prepared by the PLD method
~\cite{Kaung-Hsiung_Wu_1998,Chen2013}. 
The 248 nm KrF excimer laser was operated at a 5 Hz repetition rate with an energy density of 3 J/cm$^2$.  The laser beam, with a pulse duration of 25 ns, was focused onto the target surface through a lens, resulting in a spot size of 1 mm $\times$ 3 mm. The laser beam vaporized the YBCO or YCBCO targets, which were then deposited onto a
 one side polished $10\times10\times0.5\,\rm mm^3$-sized (100) MgO substrates at 
750 $ ^\circ$C with oxygen pressure of 0.3 Torr. The thickness of the YBCO or YCBCO thin films~$d_\mathrm{f}$ was controlled to be approximately between 110 and 140 nanometers according to the estimation by excimer laser repetition rates multiplied by the deposition duration and confirmed by scanning electron microscopy (SEM, JEOL JSM7001F). After the PLD, the underdoped YBCO thin film was prepared by post-annealing at 450$ ^\circ$C  with oxygen pressure of 6 Torr, and the overdoped YCBCO thin film was prepared by post-annealing at 450 $ ^\circ$C with oxygen pressure of 10 Torr. All the thin films were first characterized by X-ray diffraction (XRD, BrukerD8) for information about the crystallinity and impurities. The results indicate that all the samples are pure c-axis-oriented YBCO or YCBCO.

\subsection{DC resistivity}

\begin{figure}[t]
\centering
\includegraphics[width=\columnwidth]{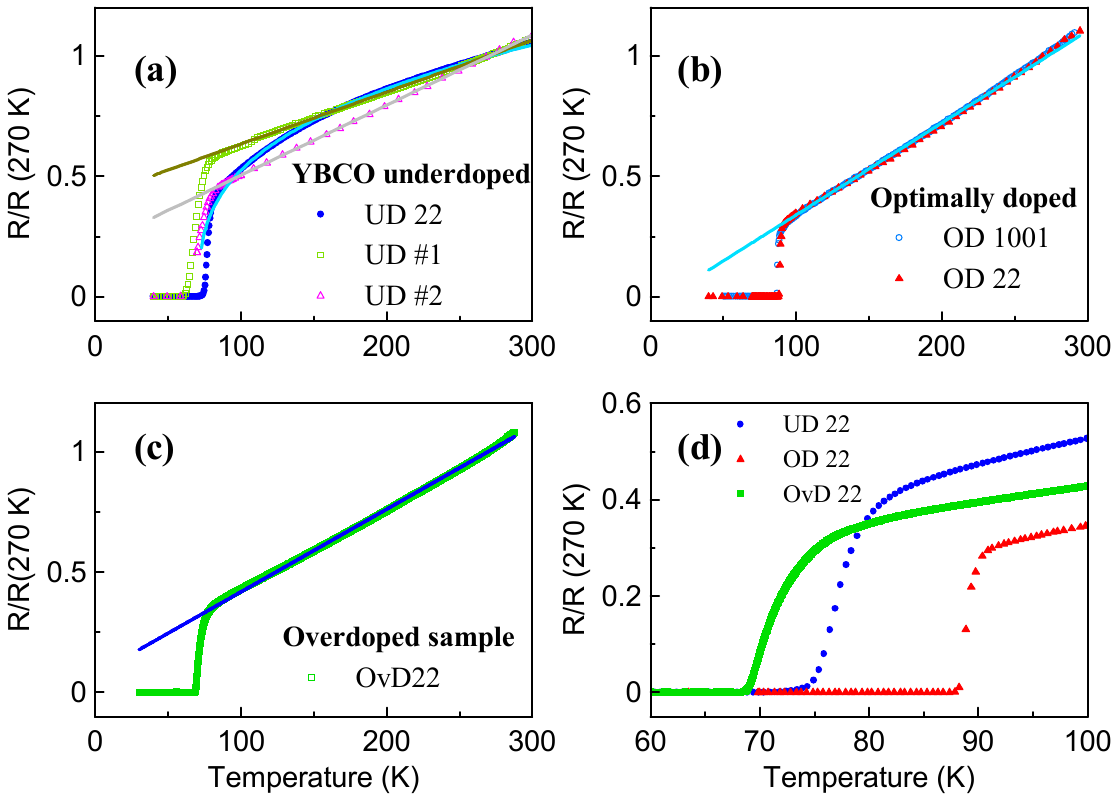}
\caption{
Temperature dependence of the resistivity for underdoped samples (panel a), optimally doped samples (panel b) and overdoped sample (panel c). Details of the resistivity around the transition are given for three selected samples in panel (d).
}
\label{R_T}
\end{figure}

 The transport measurements were applied to determine the superconductivity and transition temperatures of the films. The temperature-dependent resistivities $R(T)$ (Fig.~\ref{R_T}) were measured by the standard four-probe configuration in physical properties measurement systems (PPMS, Quantum design PPMS). The sharp transition to the superconducting state suggests the high quality of the samples. The critical temperatures $T_\mathrm{c}$ reported here were determined as  an average value of  the temperature where the resistance starts to drop and the temperature where resistivity reaches zero. The critical temperatures and the hole dopings of the samples are listed in~Table \ref{vzorky}, where UD stands for underdoped, OD for optimally doped and OvD for overdoped sample.

In all but one measurement, the DC resistivity in the normal state is very well described by a linear dependence.  The optimally doped sample OD 22 exhibits a nearly identical resistivity dependence as the sample prepared previously~\cite{Tesar2021}, even though the latter was grown on LAO and OD 22 on MgO. For the underdoped sample, we observed that the resistivity deviates from the linear behaviour.
 Phenomenologically, we found that R(T) is proportional to $(T-T_\mathrm{c})^{\beta_R}$ where the best fit gives $\beta_R = 0.364 \pm 0.002$.

It is possible to determine the hole doping from the critical temperature. Presland~\cite{Presland1991} reported that in various high-$T_\mathrm{c}$ superconductors or various stoichiometries, the hole doping can be determined from the parabolic relation $T_\mathrm{c}/T_\mathrm{c}(max) = 1 -82.6 (p-0.16)^2$, where $T_\mathrm{c}$(max) is the maximal critical temperature for the optimal doping. In our case, it is $T_\mathrm{c}$(max)=92~K for all YBaCuO samples and  $T_\mathrm{c}$(max)=85~K for the sample with Ca atoms.  However, Liang~\cite{Liang2006} observed deviations from this formula in YBCO samples and proposed a corrected dependence of the hole doping on $T_\mathrm{c}$. Here, we denote the corrected hole doping as $p^{\ast}$. 

\tabcolsep=8pt
 \begin{table}[htb]
 \caption{Resistivity measurements:} 
\label{vzorky}
\begin{tabular}{lclcccc} 
\hline
\hline
 Sample &  $d_\mathrm{f}$ &  $T_\mathrm{c}$ & $p$ & $p^{\ast}$\\
  \ &  (nm)  & (K) &   &    \\
\hline
UD \#1 &70  & 70 &.106 & .128\\
UD \#2 &100  & 70 &.106 & .128 \\
UD 22 &  140&  78  &.116  & .135  \\
OD 1001~\cite{Tesar2021} &107 &  88&.137 &  .146    \\
OD 21 & 250 &89&  .140& .147   \\
OD 22 & 140 & 89 & .140 & .147\\
OvD 22  &110  &    72& .203 & .203 \\
\hline
\hline
\end{tabular}
\end{table}

\subsection{Terahertz spectroscopy}

The experiments were performed using a custom-made THz time-domain spectrometer powered with a Ti:sapphire femtosecond laser oscillator (Coherent, Mira). For the generation of linearly polarized THz pulses we used a biased-semiconductor emitter (TeraSED, GigaOptics); for their detection, we applied the usual electro-optic sampling scheme \cite{Nahata1996} with a 1\,mm thick $\langle110\rangle$ ZnTe crystal. The measurements were realized in the transmission geometry under normal incidence in a helium flow cryostat (Optistat, Oxford Instruments).

In the experiment, three different timeforms of the electric field of the THz pulse $E(t)$ were measured, corresponding to: (i) the free aperture, (ii) the reference substrate and (iii) the superconducting film on the substrate. 
Using the echoes seen in the time profiles due to the internal reflections in the samples, the thicknesses of the reference $d_{\rm ref}$ and of the film-supporting substrate $d_{\rm sub}$ were determined, as well as their complex refractive index $\tilde{n}_{\rm sub}(\nu)$, for the details see the supplemental material~\cite{SM}.
The frequency~($\nu$) dependence of the complex transmittance $\tilde{t}(\nu)$ was evaluated as
the ratio between the Fourier transforms $E_{\rm s}(\nu)$ and $E_{\rm r}(\nu)$ of
the time profiles transmitted through the sample (film on MgO substrate) 
 and a bare MgO reference of a similar thickness.
 The complex conductivity  $\tilde\sigma(\nu)$ of the film can then be calculated from
 \cite{conventionnote}:
\begin{equation}
\label{t_formula}
\tilde{t}(\nu)=\frac{E_{\rm s}(\nu)}{E_{\rm r}(\nu)}=\frac{\left[ 
  1+\tilde{n}_{\rm
 sub}(\nu)\right] \mbox{e}^{\rm i \psi(\nu)}}{1+\tilde{n}_{\rm sub}(\nu)+Z_0 \tilde{\sigma}(\nu)d_{\mathrm{f}}},
\end{equation}
where $Z_0$ denotes the vacuum impedance, 
and $\psi(\nu)$ is the phase delay due to different optical thicknesses of the sample and
the reference. 
Since the same material is used for the substrate and the reference, we can write
\begin{equation}
  \psi(\nu) = \frac{2\pi\nu}{c}\left[(\tilde{n}(\nu) - 1)d_{\rm f}+\tilde{n}_{\rm sub}(\nu)(d_{\rm sub}-d_{\rm ref})\right],
\end{equation}
 where $c$ is the light velocity and $\tilde{n}$ is the complex refractive index of the film, and $d_{\rm sub}$ is  the thickness of the film-supporting substrate. 
 The first term in the square brackets stems from  the propagation in  the film, and the second one reflects the different thicknesses of the sample and reference substrates.

The complex conductivity of the film $\tilde{\sigma}(\nu)$ is linked to its refractive index via the permittivity by the relationships  $\tilde{\varepsilon}(\nu) = \tilde{n}(\nu)^2$ and
$\tilde{\sigma}(\nu)= -{2\pi \rm i}\nu\varepsilon_0 \tilde{\varepsilon}(\nu)$, where $\varepsilon_0$
denotes the vacuum permittivity.

\begin{figure}[t]
\centering
\includegraphics[width=0.87\columnwidth]{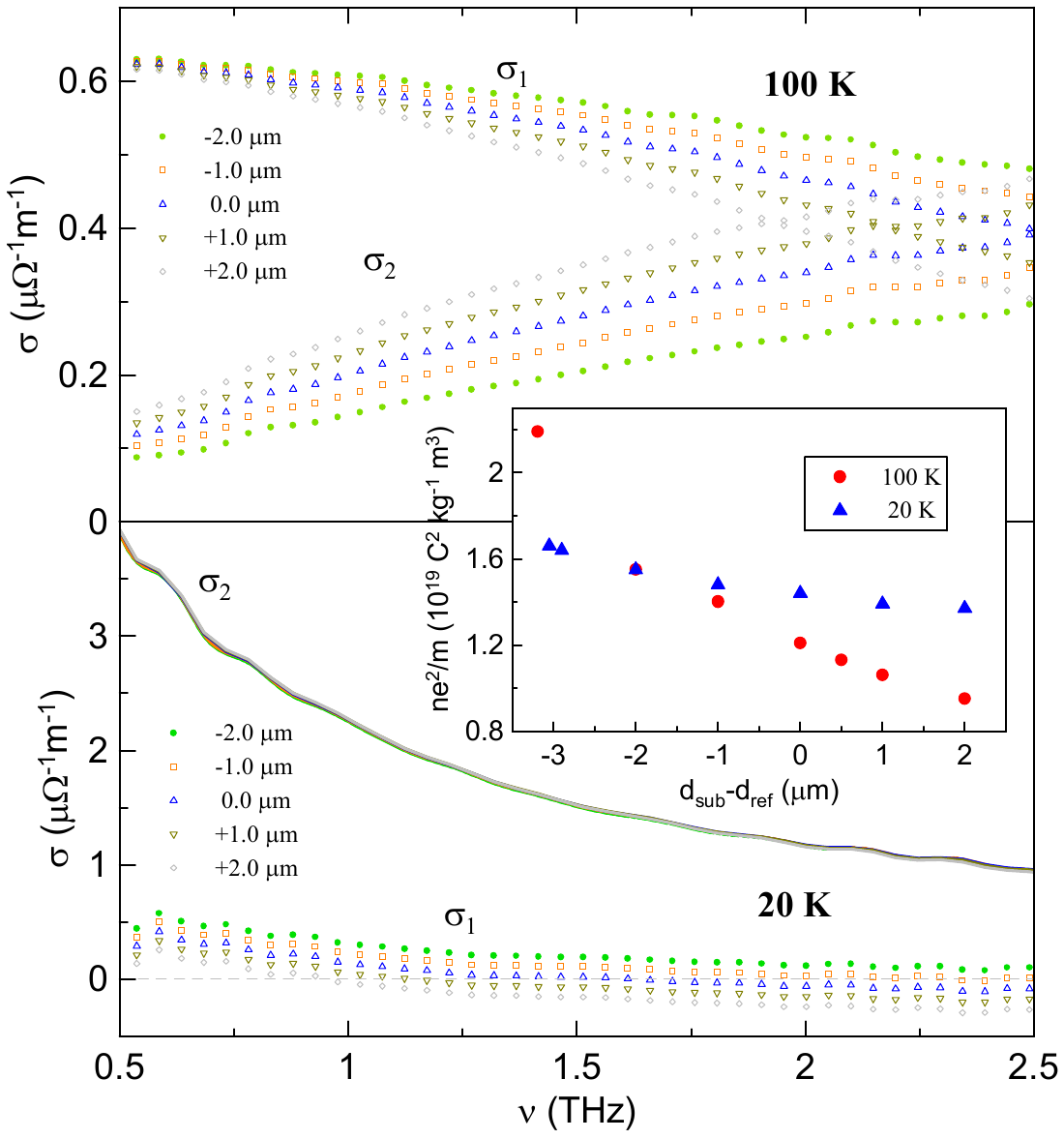}
\caption{
Complex conductivity $\tilde{\sigma}(\nu)$ of the optimally doped sample OD 21 in the normal state at 100 K (upper panel) and in the superconducting state at 20 K (bottom panel) evaluated for various thickness differences ($d_{\rm sub}-d_{\rm ref}$).
 In the inset, we evaluated $ne^2/m$ assuming different values of ($d_{\rm sub}-d_{\rm ref}$) and found the agreement for 20~K and 100~K measurements in case of $d_{\rm sub}-d_{\rm ref}$ = -2 $\mu$m  (highlighted by {\color{green}$\bullet$}) thus this value was used to evaluate the complex conductivity.
}
\label{phase}
\end{figure}

The accuracy of the measurement of the complex conductivity is determined by the plan-parallelity of the MgO slabs, both the reference and the film-supporting one. The phase measurement is particularly sensitive to the thickness difference ($d_{\rm sub}-d_{\rm ref}$), as even a small error in it may cause large errors in the determination of the 
conductivity, see figure~\ref{phase}. While the imaginary part of the conductivity $\sigma_2(\nu)$ at low temperature is rather insensitive to the phase determination,  $\sigma_1(\nu)$ is quite sensitive to it and an incorrect thickness difference leads to a constant offset of $\sigma_1$. In the normal state, the thicknesss difference change the slope of both $\sigma_1(\omega)$ and $\sigma_2(\omega)$. The thickness difference was estimated from the time delay of the pulse reflected in the reference and in the substrate and it was checked that, within the experimental accuracy, the quantity $ne^2/m$ is the same in the superconducting and in the normal state, see the inset in Fig.~\ref{phase}.

\begin{figure*}[hbt]
\includegraphics[width=2\columnwidth]{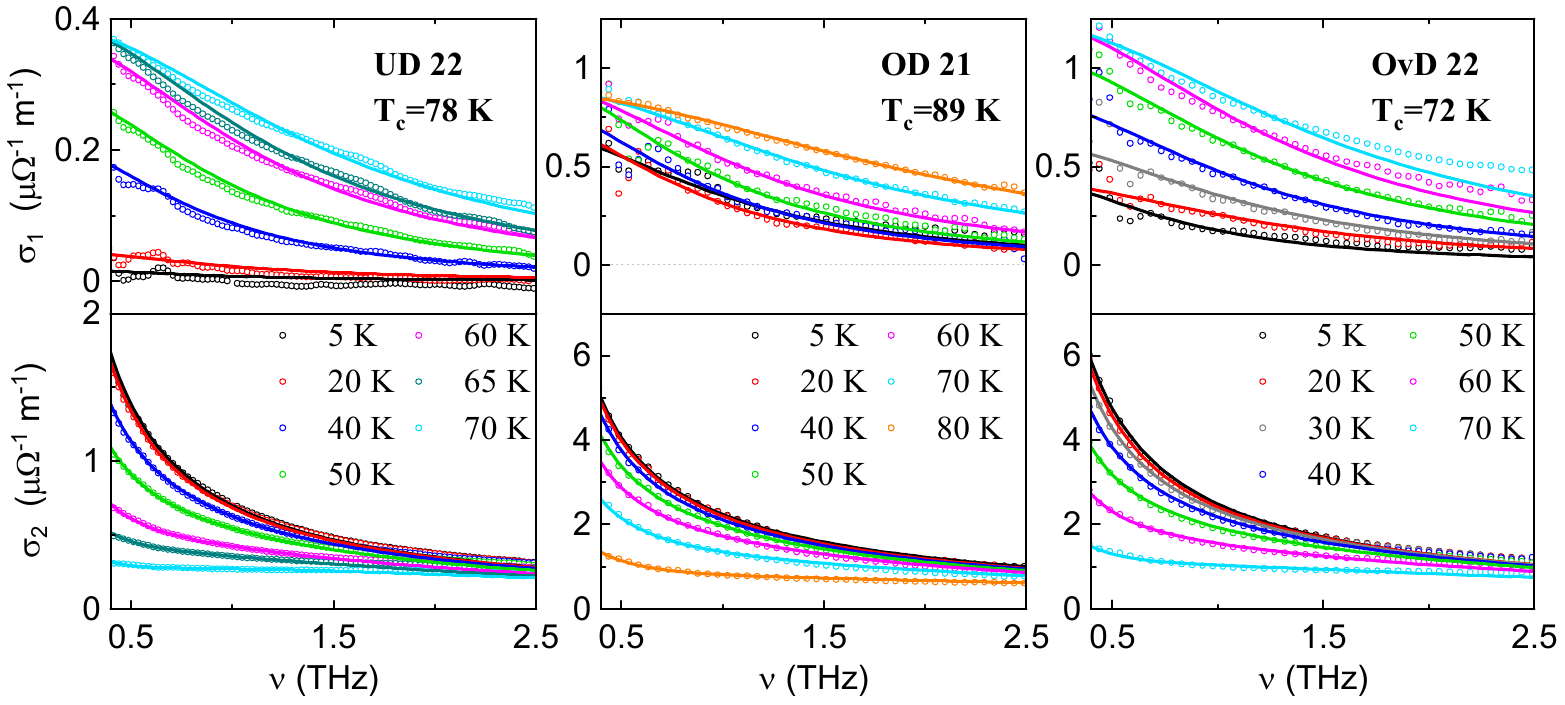}
\includegraphics[width=2\columnwidth]{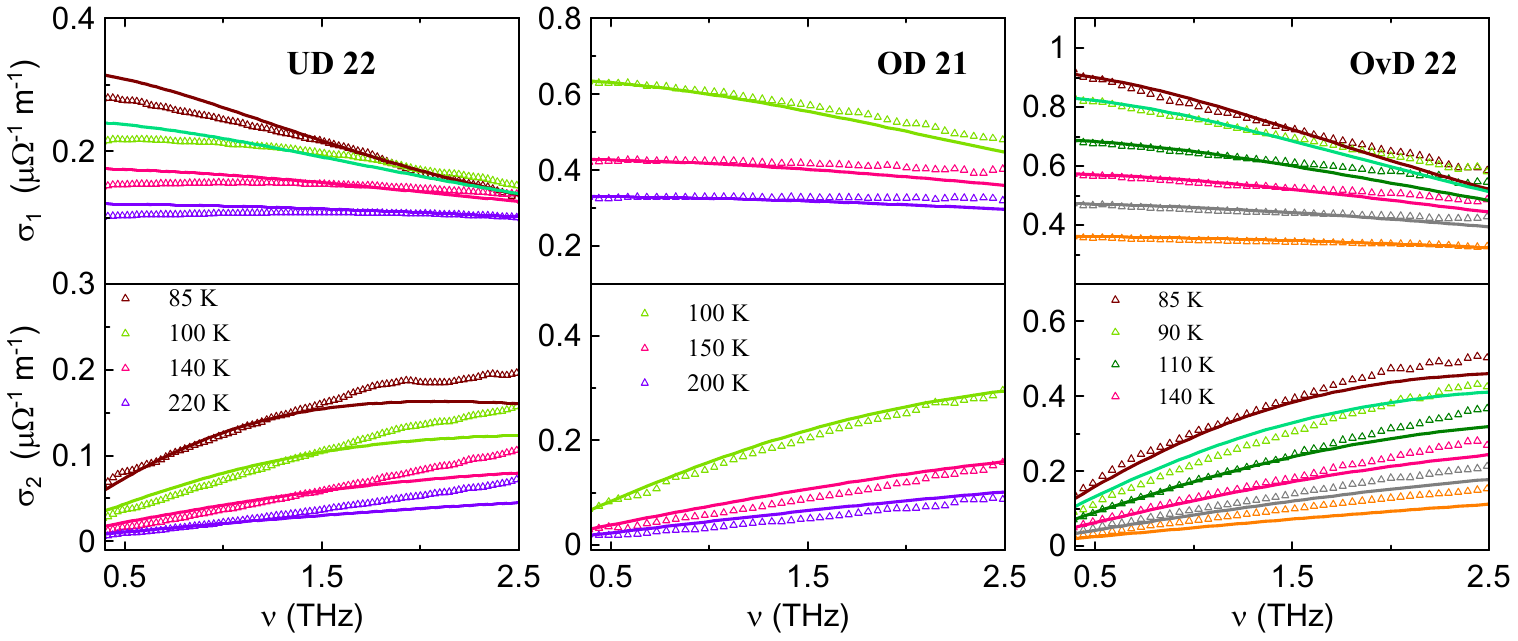}
\caption{Real  and imaginary  parts of the complex conductivity of underdoped UD22 (left), nearly optimally doped OD 21 (middle) and overdoped OvD22 (right) samples for $T<T_\mathrm{c}$ (top) and for $T>T_\mathrm{c}$ (bottom). Points - experimental data, curves - fit by the two-fluid model (eq.~\ref{2f_model}) for $T<T_\mathrm{c}$  and by the Drude model for $T>T_\mathrm{c}$.}
\label{s_sc_all}
\end{figure*}

\section{Results and discussion}

 Here, we present the terahertz conductivity up to  2.5 THz for three different dopings of YBa$_2$Cu$_3$O$_{7-\delta}$  in Fig.~\ref{s_sc_all} for the superconducting state (upper graphs) and for the normal state (bottom graphs): the optimally doped OD 21 (middle), the underdoped UD 22 (left) and the overdoped OvD 22 (right) cases. We underline that not all experimentally acquired data are shown in the graphs for clarity.

In the superconducting state, the two-fluid model describes the complex conductivity quite well for all observed samples. In the fitting procedure, we used measured data in the frequency interval from 0.4 to 1.6 THz where we have the strongest signal thus the data are the most reliable. The obtained theoretical curves are shown even beyond their fitting range.  Below $T_c$, the quantities $n_\mathrm{s}e^2/m$,  $n_\mathrm{n}e^2/m$  and $\tau$ were free parameters and the total electron concentration was not fixed, see the table~\ref{vzorkyII}. Above $T_c$, the Drude model was used with the total electron concentration fixed to the value determined below $T_c$, see Fig.~\ref{s_sc_all} thus the scattering time $\tau$ is the only free parameters in these fits. 
 The real part $\sigma_1$ is well described by the Drude response. 
In the normal state, in the fitting frequency interval, $\sigma_1$ dominates over $\sigma_2$, in contrast with the low-temperature limit below $T_\mathrm{c}$.  Thus $\sigma_1$ is determined more reliably here. In optimally doped and overdoped cases, the Drude model describes the data both qualitatively and quantitatively reasonably well. However, for the underdoped sample, the experimental data starts to deviate from the Drude response, which is not surprising given the unusual properties of YBaCuO in the pseudogap regime. We speculate that for underdoped samples with lower $T_\mathrm{c}$, the discrepancy will even grow.

\begin{figure}[bt]
\includegraphics[width=\columnwidth]{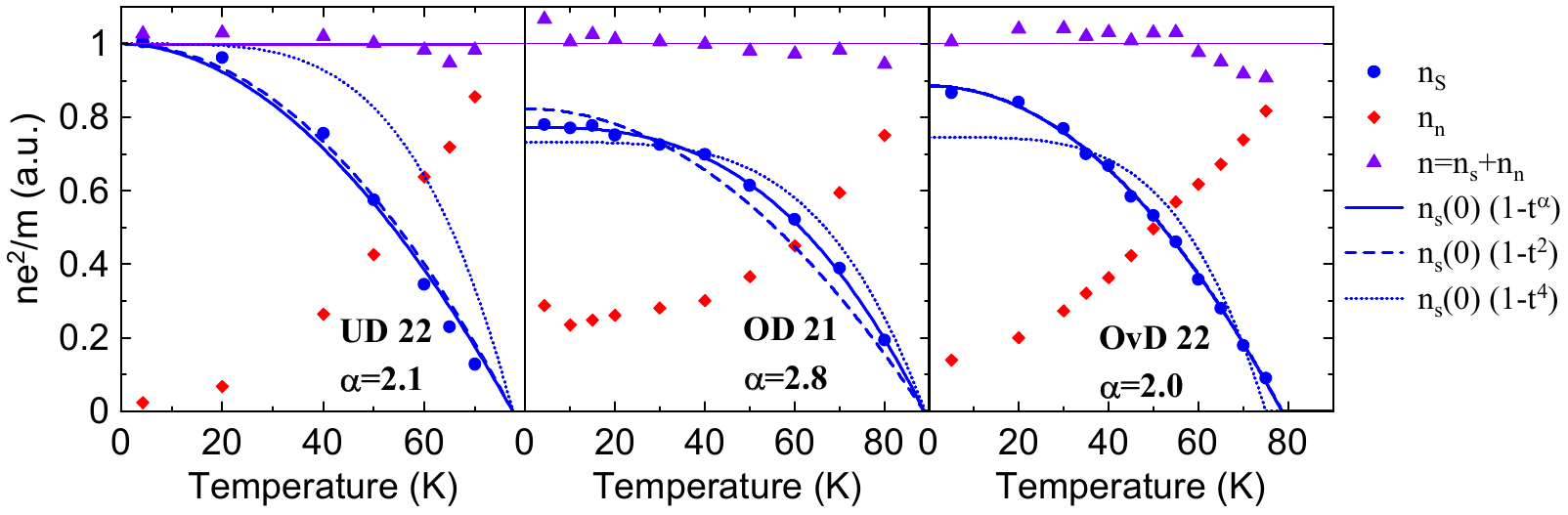}
\caption{ Temperature dependence of the superfluid (circles), the normal fluid (diamond) and the total volume fraction (triangles)  of underdoped UD22 (left),  nearly optimally doped OD 21 (middle) and overdoped OvD22 (right) samples. $ne^2/m$ is fitted by a constant (see Table~\ref{vzorkyII}) and the data are normalized on its value. The temperature dependence of the superfluid fraction is described by $1-(T/T_c)^{\alpha}$ (solid line), $1-(T/T_c)^2$ (dashed line) and $1-(T/T_c)^4$ (dotted line).
}
\label{fs_all}
\end{figure}

First, let's analyze the contributions from the superfluid and from the normal fluid to the conductivity using the two fluid model, see figure~\ref{fs_all}. $ne^2/m$ is well described by a constant fit for all temperatures in all three studied samples, see the values in table~\ref{vzorkyII}. It is convenient to use a normalization by dividing by $ne^2/m$. Assuming that the effective mass does not change in the superconducting state, we can write the superfluid fraction as $f_\mathrm{s}(T)=n_\mathrm{s}(T)/n$ and the normal fluid fraction as $f_\mathrm{n}(T)=n_\mathrm{n}(T)/n$. One of the most important questions is whether all electrons condensate to superfluid fraction as it was observed by some works ~\cite{Brorson1996,Buhleier1994,Shibauchi1992,Ludwig1996} or if there is a finite fraction of normal fluid as reported in other papers~\cite{Frenkel1996,Pimenov1999,Pimenov2000,Waldram1997}.
Frenkel~\cite{Frenkel1996} observed $f_n=0.35$ at $T=20$~K in a YBaCuO thin film deposited on LAO with $T_c=90$~K. Below 10~K, Pimenov observed $f_n \approx 1/4 $ in a thin YBCO film with $T_c=89.5$ K~\cite{Pimenov1999} and a similar value of $f_n \approx 1/5 $ in an underdoped film with $T_c=56.5$~K~\cite{Pimenov2000}. Waldram~\cite{Waldram1997} argued that even good quality thin films may contain 10\% of normal electrons at $T=0$~K. Our results, see figure~\ref{fs_all}, suggest that all electron condensate in the underdoped film and that there is a finite normal fluid fraction in the nearly optimally doped ($f_n=0.25\pm0.05$ at 5 K) and in the overdoped film ($f_n=0.14\pm0.01$ at 5 K), see Table~\ref{vzorkyII}. Mahmood~\cite{Mahmood2019} observed an even more drastic fraction of uncondensed electron in overdoped La$_{2-x}$Sr$_{x}$CuO$_4$ samples with $T_c$ in the interval between 30 K and 5 K and corresponding $f_n$ from 0.5 to 0.9. Lee-Hone \emph{et al}~\cite{Lee-Hone2018} argued that the fraction of uncondensed electron scales with the disorder.

The experimentally determined superfluid fraction $f_s(T)$ decreases towards the critical temperature and is well described by the formula $f_s=f_s(0) (1-t^{\alpha})$ following equation~\ref{lambda_T_dep}. The values of the exponents $\alpha$ are listed in table~\ref{vzorkyII}. The values of $\alpha$  for the underdoped and overdoped samples are clustered around 2, corresponding to the quadratic dependence which is predicted for d-wave superconductors with strong impurity scattering~\cite{Prohammer1991}. 
The nearly optimally doped sample differs having the value $\alpha=2.8$.
 We did not observe the pronounced linear dependence which is predicted for the
case of d-wave pairing, contrary to Bozovic's findings in La$_{2–x}$Sr$_x$CuO$_4$  films in a broad range of oxygen doping~\cite{Bozovic2018}. The transition from $\alpha=1$ (the linear dependence of $f_s$) to the higher values of $\alpha$ can be explained by introducing disorder, see the figure 5 in  the reference~\cite{Lee-Hone2020}.

The penetration depth in the zero-temperature limit $\lambda_L(0)$ was determined from the fitted value of $n_se^2/m$ at the lowest measured temperature for all samples, see table~\ref{vzorkyII}. 
Previous studies have reported that the zero-temperature penetration depth values in optimally doped YBaCuO samples typically fall within the range of 0.14 - 0.18~$\mu$m~\cite{Bonn_Hardy_V,Ludwig1996,Nuss1991,Bonn1993,Basov1994,Sonier1994}.  
Vaulchier~\cite{Vaulchier1996} reported the value of 0.16 $\mu$m for the sample with $T_c=92$~K and the value of 0.34~$\mu$m for the sample with a slightly lower critical temperature $T_c=86$~K. 
An even higher value ($\lambda_L(0)=0.38\,\mu$m) was found by Tsai~\cite{Tsai2003} in a high-resistivity YBCO thin film with $T_c=84$~K. For the optimally doped sample, we found $\lambda_L(0)=0.28\,\mu$m, which is higher than for most of the reported values but still lower than those of Vaulchier~\cite{Vaulchier1996} or Tsai~\cite{Tsai2003}. Quite a high value of $\lambda_L(0)$ suggests an extrinsic contribution~\cite{Vaulchier1996}.
Hylton~\cite{Hylton1988} suggested a phenomenological formula $\lambda^2=\lambda_{intr}^2+\lambda_{extr}^2$ accounting for intrinsic and extrinsic contributions,  such as for example those due to weak links.
 Ludwig~\cite{Ludwig1996} found values of 0.18~$\mu$m in their optimally doped sample ($T_c=90$~K) and twice larger values for their two underdoped samples 0.36~$\mu$m and 0.38~$\mu$m with critical temperatures 65~K and 57~K. Our underdoped sample exhibits a slightly higher value $\lambda_L(0)=0.43\,\mu$m while it has a higher $T_c=78$~K. Finally, the value of the London penetration depth $\lambda_L(0)$ for the overdoped sample Y$_{0.7}$Ca$_{0.3}$Ba$_2$Cu$_3$O$_{7-\delta}$ is $=0.23\,\mu$m, which is comparable to the value for our nearly optimally doped sample and higher than the one estimated by Bachar ($\lambda_L(0)=(0.18-0.20)\,\mu$m)~\cite{Bachar2011} for a similar sample.

 \begin{table*}[hbt]
 \caption{Parameters of the investigated samples.\\
} 
\label{vzorkyII}
\begin{tabular}{lccccccc}
\hline
 Sample & $T_c$ & $\alpha$  & $ne^2/m$                    &  $f_n$ (5~K)                              &  $\lambda_L(0)$  & $1/\tau_0$ & $\lambda$\\
            & K &                &     $10^{19}$ C$^2 $kg$^{-1}$m$^{3}$  &   & $\mu$m  & ps$^{-1}$ &\\ 
\hline
UD 22 & 78 &$2.1\pm 0.1 $ & $0.43\pm 0.01$ & 0.02 $\pm$ 0.01 & 0.43&  4.5 $\pm$ 0.9 & 0.173\\
OD 21 &  89  & $2.8 \pm 0.1 $  &$1.53 \pm 0.02$ &  0.25 $\pm$ 0.05& 0.28&     2.4 $\pm$ 0.5    & 0.225 \\
OvD 22 & 72 & $2.0 \pm 0.1$ &$1.65 \pm 0.02$&  0.14 $\pm$ 0.01 &0.23&    6.3 $\pm$ 0.2    &  0.191  \\
\hline
\end{tabular}
\end{table*}

\begin{figure}[hbt]
\includegraphics[width=\columnwidth]{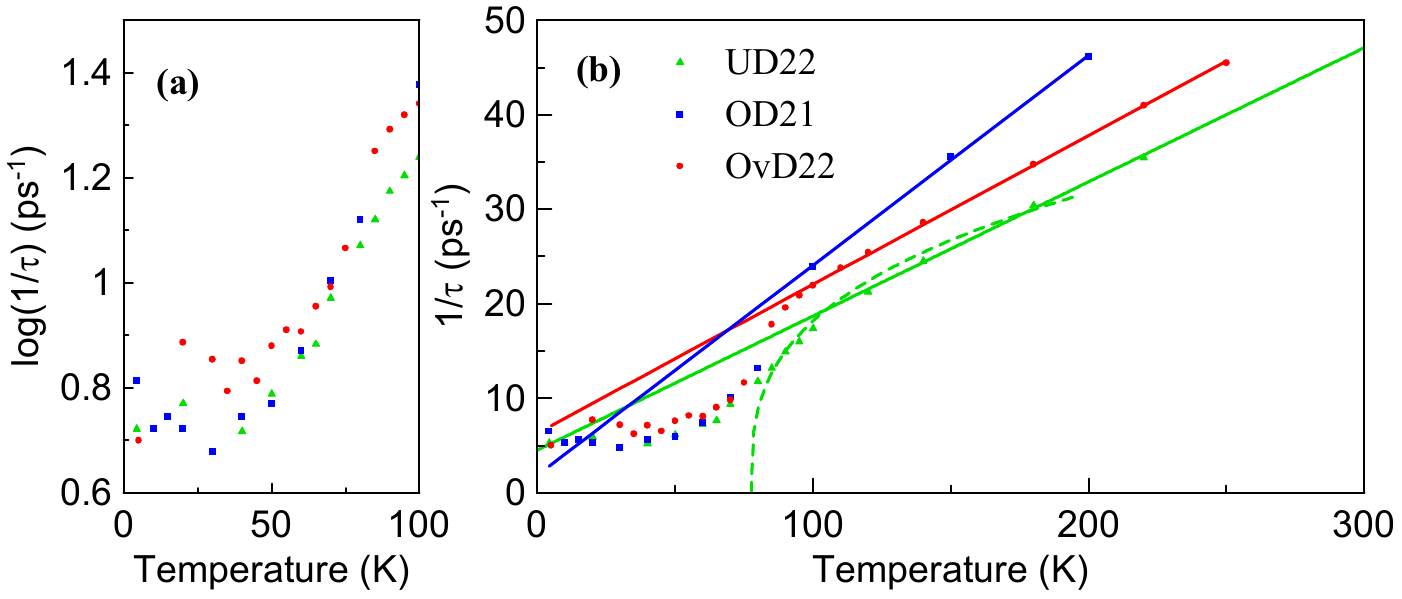}
\caption{Scattering rate $1/\tau$ for underdoped UD22 (green triangles), optimally doped OD21 (blue squares) and overdoped OvD22 (red circles) samples. The solid curves corresponds to the fits described in detail in the text and their extrapolation beyond the fitting range. (a) Logarithmic scale of the scattering rate $1/\tau$ below 100 K, (b) Linear scale of $1/\tau$ for the whole temperature range.}
\label{1_tau}
\end{figure}

Below $T_\mathrm{c}$, the scattering rate $1/\tau$ was determined from the Drude response function, since it accounts for the normal fluid contribution. 
The temperature dependence of the scattering rate is plotted in Fig.~\ref{1_tau} for all cases. 
 Above the critical temperature,  $1/\tau$ is well described by a linear dependence in agreement with the linear behaviour of the DC resistivity, see graph~\ref{R_T}. The nearly optimally doped and overdoped samples exhibit a linear dependence from $T_c$, while the underdoped sample exhibits a linear behaviour only above approximately 120~K. Between $T_c$ and 120~K, $1/\tau$  is proportional to $(T-T_\mathrm{c})^{\beta_{\tau}}$, where we found the best agreement for $\beta_{\tau}=(0.33 \pm 0.02)$, which agrees well with the DC resistivity exponent $\beta_R$. 
The linear dependence is described by the formula~\cite{Liu2006}:
\begin{equation}
\frac{\hbar}{\tau}=2\pi\lambda k_B T+\frac{\hbar}{\tau_0},
\label{Liu_eq}
\end{equation}
where $\lambda$ is a dimensionless electron-phonon coupling parameter, $\hbar$ is the reduced Planck constant, $k_B$ is the Boltzmann constant and $1/\tau_0$ is the zero-temperature intercept,
the residual scattering rate. The extracted values of $\lambda$ and $1/\tau_0$ are listed in Table~\ref{vzorkyII}. While the values of $1/\tau_0$ falls within the range reported by Liu~\cite{Liu2006}, the values of $\lambda$ are approximately half of its typical reported value of 0.37~\cite{Liu2006}.

Below $T_c$ the scattering rate exponentially falls down  and saturates for temperatures below 50~K for all samples, see Fig~\ref{1_tau}a.
It is worth noting that in the superconducting state, the scattering rates  for all measured samples in the temperature interval below $T_c$ align closely with each other. 
A steeper initial drop below $T_c$ over two orders of magnitude was
 reported in YBaCuO crystals~\cite{Bonn1994} and thin films~\cite{Gao1993,Buhleier1994,Brorson1996}. While some works~\cite{Bonn1994,Gao1993} observed  $1/\tau(T)$ to saturate below 
40~K, in others~\cite{Buhleier1994,Brorson1996} no sign of saturation is visible even at their lowest measured temperature about 30~K.

\begin{figure}[thb]
\includegraphics[width=\columnwidth]{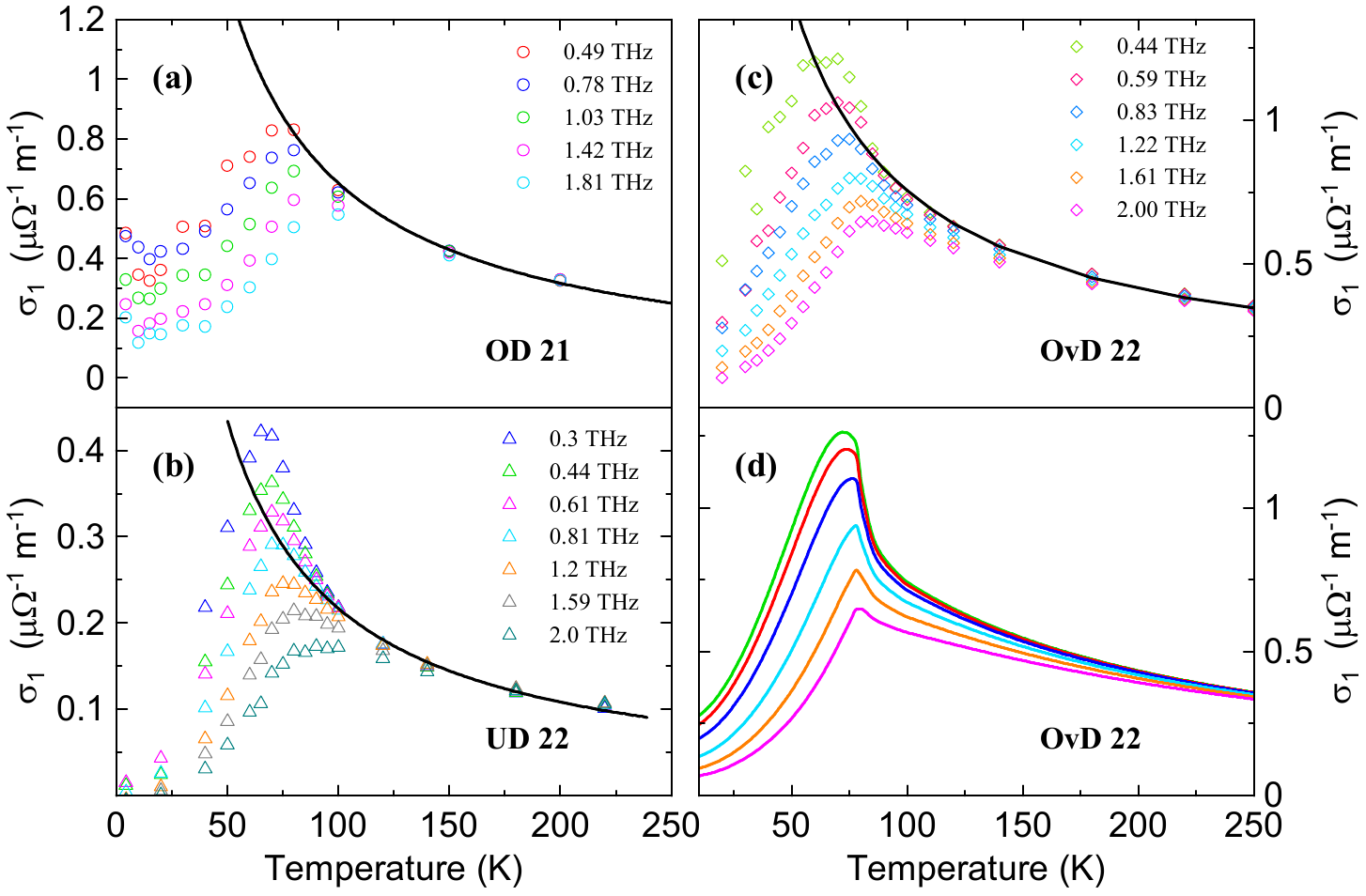}
\caption{
Temperature dependence of the real part of the conductivity  $\sigma_1(T)$  for different frequencies for the three cases of doping: (a) nearly optimally doped OD21, (b) underdoped OD22 and (c) overdoped OvD22 samples. The peaks shift only marginally with the frequency.  In the normal state, the conductivity is inversely proportional to the temperature, which is shown by a solid line for one frequency in each case. (d) A reconstruction of the temperature dependence of the conductivity is shown for the optimally doped sample OvD 22.
}
\label{s1_peak}
\end{figure}

We extracted the temperature dependence of the real part of the complex conductivity $\sigma_1(T)$ at given frequencies from the spectra presented in Fig~\ref{s_sc_all}. A peak can be observed for each doping and its exact shape depends on the quality of the sample (Fig.~\ref{s1_peak}).
Similar peaks in $\sigma_1(T)$  were reported in previous studies in YBCO samples~\cite{Nuss1991,Ludwig1996,Pimenov1999,Pimenov2000}. At low frequencies, the real part of the conductivity $\sigma_1$ is determined less reliably due to the dominating $1/\omega$ frequency dependence of the imaginary part $\sigma_2$, thus the data below 0.3 THz show some unphysical fluctuations.
 From room temperature down to the critical temperature, $\sigma_1(T)$ grows following  the $1/T$ dependence of the scattering time. Below $T_\mathrm{c}$, $\sigma_1$ steeply increases, reaches its maximum and decreases as the normal fluid fraction sharply falls. The peak position depends only slightly on the frequency.  This frequency dependence suggests that it does not correspond to the coherence peak predicted by the BCS theory, as it was discussed in  Frenkel paper~\cite{Frenkel1996}.

The analysis of the spectra suggests that this peak is the product of the competition between $\tau(T)$ and $f_\mathrm{s}(T)$.
The scattering time and the normal fluid fraction, obtained from the fits in Fig.~\ref{1_tau}  and \ref{fs_all}, were interpolated and these values were used in  the equation~\ref{2f_model}
to reconstruct $\sigma_1(T)$.
The model qualitatively and semi-quantitatively describes all the data for all frequencies (Fig.~\ref{s1_peak}d).
\\

\section{Summary and conclusion}

We prepared a high-quality overdoped thin Y$_{0.7}$Ca$_{0.3}$Ba$_2$Cu$_3$O$_{7-\delta}$ (YCBCO) film and underdoped and optimally doped thin films of its maternal compound YBa$_2$Cu$_3$O$_{7-\delta}$ and analyzed their properties in the low frequency regime using THz spectroscopy. All these samples were prepared by pulsed laser deposition on MgO substrate at similar operating conditions except for different oxygen pressure during the deposition. The DC resistivity measurements show sharp superconducting transitions indicating the high quality of the samples.

Above the critical temperature, the Drude model describes well the real part of the conductivity $\sigma_1(\nu)$,  but for the overdoped and the underdoped samples, we observed some deviations in the imaginary part of conductivity $\sigma_2(\nu)$. 
Below the critical temperature, the two-fluid model successfully describes the response of all samples. 
We used the form of the two-fluid model which allows for incomplete condensation even in the low-temperature limit. 
At 5~K, all charge carrier condensate into superfluid in the underdoped sample, whereas a significant normal fraction remains in the overdoped and the optimally doped samples. It was suggested that except for the ultraclean samples, the disorder depletes the superfluid fraction~\cite{Lee-Hone2018}.
For the underdoped and the overdoped films, the superfluid volume fraction $f_\mathrm{s}(T)$  follows a quadratic temperature dependence  which hints a dirty d-wave superconductor.

For all samples, at high temperatures,  we observed a linear temperature dependence of the scattering rates $1/\tau(T)$.
Below $T_c$, they collapse on each other as they decrease exponentially until they saturates below 50~K. 
 We found a peak in $\sigma_1(T)$ not only in the underdoped and the optimally doped samples, in agreement with previous  studies~\cite{Nuss1991,Ludwig1996,Pimenov1999,Frenkel1996}, but the same feature is also present in the overdoped sample.  This peak is the result of the competition of the temperature dependence of the scattering rate $1/\tau(T)$ and the normal state fraction $1-f_\mathrm{s}(T)$.
 
 Altogether, the behaviour of the overdoped sample is
quite similar to that of the optimally doped one, whereas the underdoped sample behaves qualitatively slightly different.

\section{Acknowledgment}
We gratefully acknowledge helpful conversations with P. Lipavsk\'{y}.
We acknowledge the financial support by the Czech Science Foundation (Project 21-11089S), 
by the Ministry of Education, Youth and Sports within the LU - INTER-EXCELLENCE II program  (Project No. LUC24098),
and the National Science and Technology Council of Taiwan (Project No. 110-2119-M-A49-002-MBK)

\bibliography{YBCO.bib}

\begin{thebibliography}{37}
\bibitem [{con()}]{conventionnote}%
  \BibitemOpen
  \href@noop {} {}\bibinfo {howpublished} {We are using the convention
  $E(t)=E_0 \exp (-i\omega t)$ which implies the following forms of the complex
  quantities: conductivity $\tilde{\sigma}=\sigma_1+i \sigma_2$, permittivity
  $\tilde{\varepsilon}=\varepsilon_1+i\varepsilon_2$, and refractive index
  $\tilde{n}=n+i\kappa$.}\BibitemShut {Stop}%
\end{thebibliography}

\end{document}


\title{Supplemental material: Two-fluid model analysis of the terahertz conductivity of YBaCuO samples: optimally doped, underdoped and overdoped cases}

\author{M. \v{S}indler$^1$}
\author{Wen-Yen~Tzeng$^2$}
\author{Chih-Wei~Luo$^{3-5}$}
\author{Jiunn-Yuan Lin$^5$}
\author{Christelle~Kadlec$^1$}

\affiliation{$^1$Institute of Physics, Academy of Sciences of the Czech Republic, Na Slovance 1999/2, 18200 Prague 8, Czech Republic}
\affiliation{$^2$Department of Electronic Engineering, National Formosa University, Yunlin 632, Taiwan}
\affiliation{$^3$Department of Electrophysics, National Yang Ming Chiao Tung University, Hsinchu 30010, Taiwan}
\affiliation{$^4$National Synchrotron Radiation Research Center, Hsinchu, 30076, Taiwan}
\affiliation{$^5$Institute of Physics, National Yang Ming Chiao Tung University, Hsinchu 30010, Taiwan}

\date{\today}

\maketitle
In this supplement, we describe how we proceed with time domain terahertz spectroscopy measurements of thin superconducting films, so that we can properly evaluate their THz conductivity. Furthermore, we provide a detailed analysis of the low temperature conductivity using the two fluid model.

 \begin{figure}[h]
 \centering
 \includegraphics[width=0.8\columnwidth]{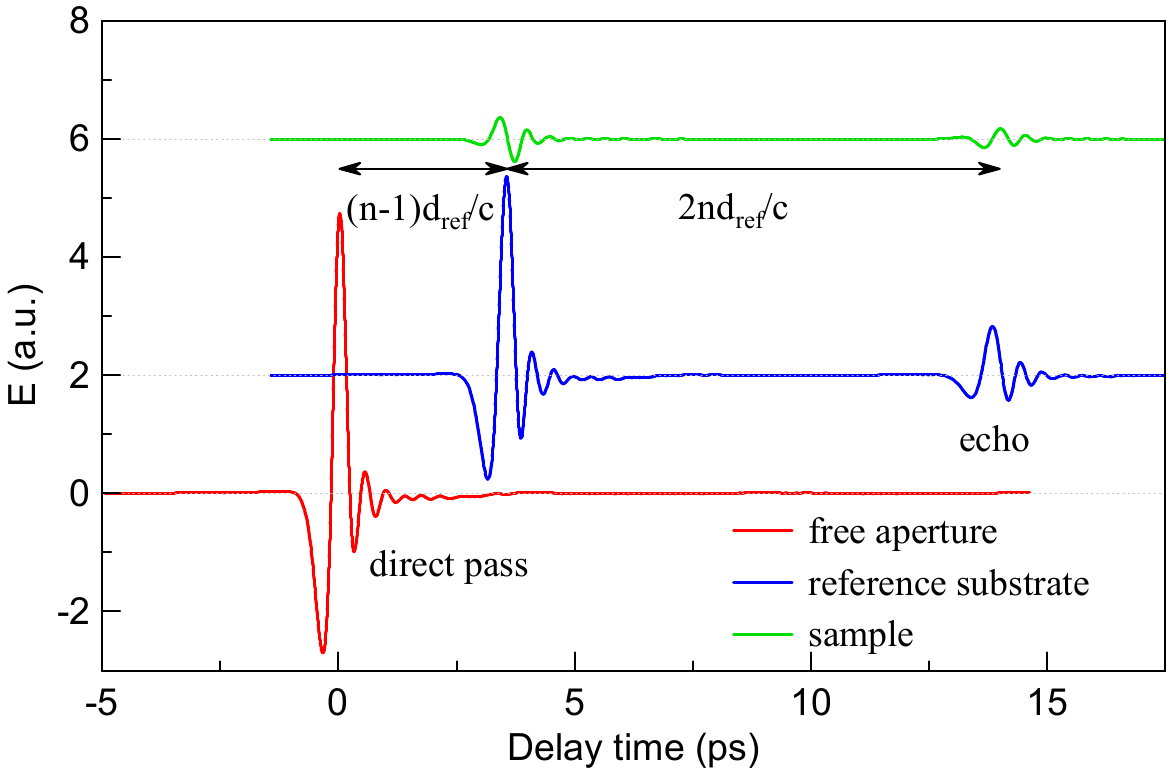}
 \caption{
Measured timeforms of the electric field of the THz pulse through a free aperture (6-mm diameter hole), the 0.5 mm thick reference MgO substrate and the underdoped YBCO sample UD22. The timeforms are vertically shifted in the graph for convenience.
 }
 
 \label{fig:timeforms}
 \end{figure}

\begin{figure}[h]
 \centering
 \includegraphics[width=0.8\columnwidth]{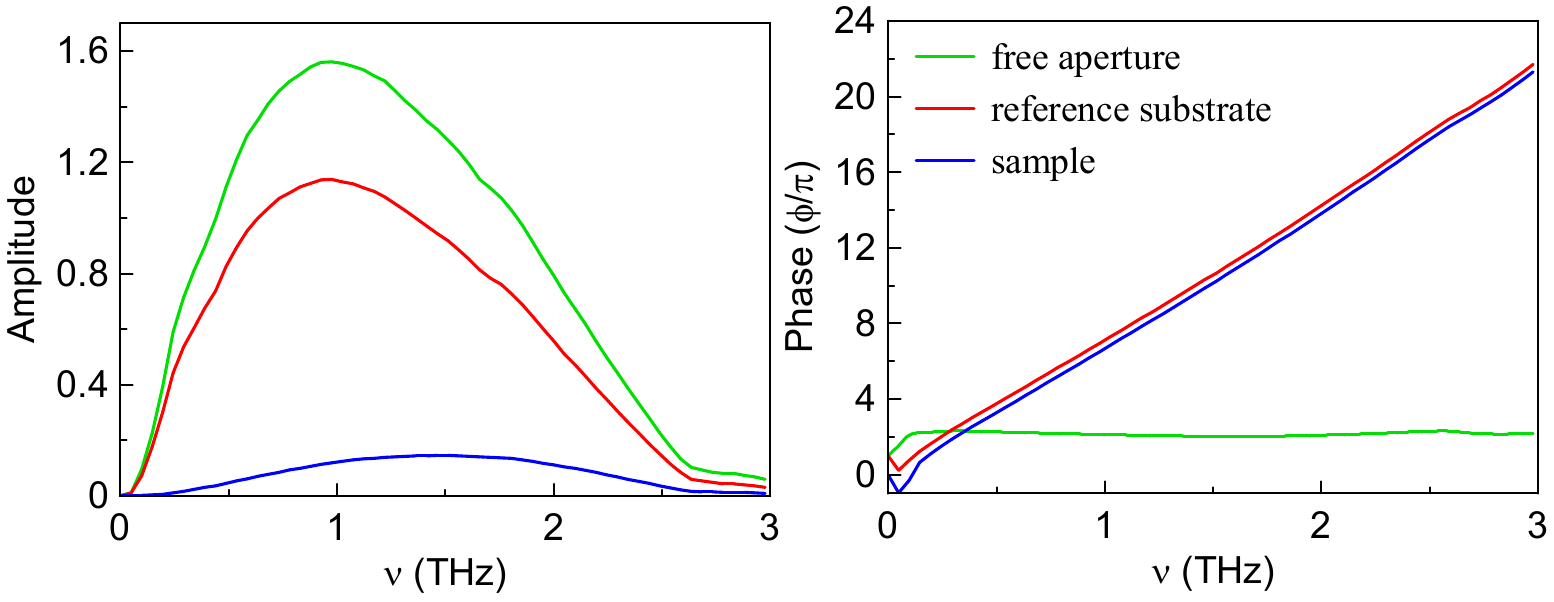}

 \caption{
Fourier transforms of the direct passes of the timeforms. We show the amplitude (left) and the phase (right).
 }
 
 \label{fig:FT}
 \end{figure}

Let's describe our experimental approach step by step. We have chosen to present the data for the underdoped sample UD22, which is the most transparent one, and thus exhibits the strongest transmission of our set of samples.
During the experiment, three different timeforms $E(t)$  of the electric field of the THz pulse are registered (see Fig~\ref{fig:timeforms}), corresponding to: (i) the free aperture (red line), (ii) the reference substrate (blue line) and (iii) the superconducting film on substrate (green line). 
Since the electric field is given in arbitrary units, a reference measurement is necessary.
Consequently, we first measure the electric field of the THz pulse transmitted through an empty aperture, then the one transmitted through a 0.5-mm thick MgO substrate. The amplitude of the signal is lower, as a consequence of the reflection at the interface and of the absorption (almost negligible for MgO). In the analysis, it is convenient to split the timeform $E(t)$ into direct pass and echo signals. In this blue curve, the first peak centered at 3 ps represents the direct pass through the substrate, and it is shifted by $(n_{sub}-1) d_{ref}/c$ compare to the free aperture, where $n_{sub}$ is the refractive index of MgO, $d_{ref}$ the thickness of the reference substrate and $c$ the speed of light. The second peak, the so-called echo, originates from the first internal reflection inside the substrate. The amplitude is further reduced by the two additional internal reflections and the signal is shifted by $2n_{sub}d_{ref}/c$. 
 The sample, consisting in the superconducting thin film deposited on a MgO substrate with a thickness similar to that of the reference substrate, is measured the same way. In this case, the internal reflections in the ultrathin film are included in the pulses and we can only distinguish the reflected pulses due to the thick MgO substrate. The difference in the  time delay for the direct pass and the echo enables us to determine the optical thickness of the  substrates, and mainly their difference.
The direct pass is analyzed by Fourier transform (FT), see Fig~\ref{fig:FT}. First, we determine the optical properties of the MgO reference substrate, especially its refractive index, from the ratio of the FT of the electric field of the reference substrate compare to that of the empty aperture $E_{r}(\nu)/E_{0}(\nu)$. Secondly, the ratio of the FT of the electric field of the sample compare to that of the reference substrate $E_{s}(\nu)/E_{r}(\nu)$ is inserted into equation (4) of the main text. The refractive index of the thin film can then be evaluated using equation (5), and finally its complex conductivity.

 \begin{figure}[h]
 \centering
 \includegraphics[width=0.8\columnwidth]{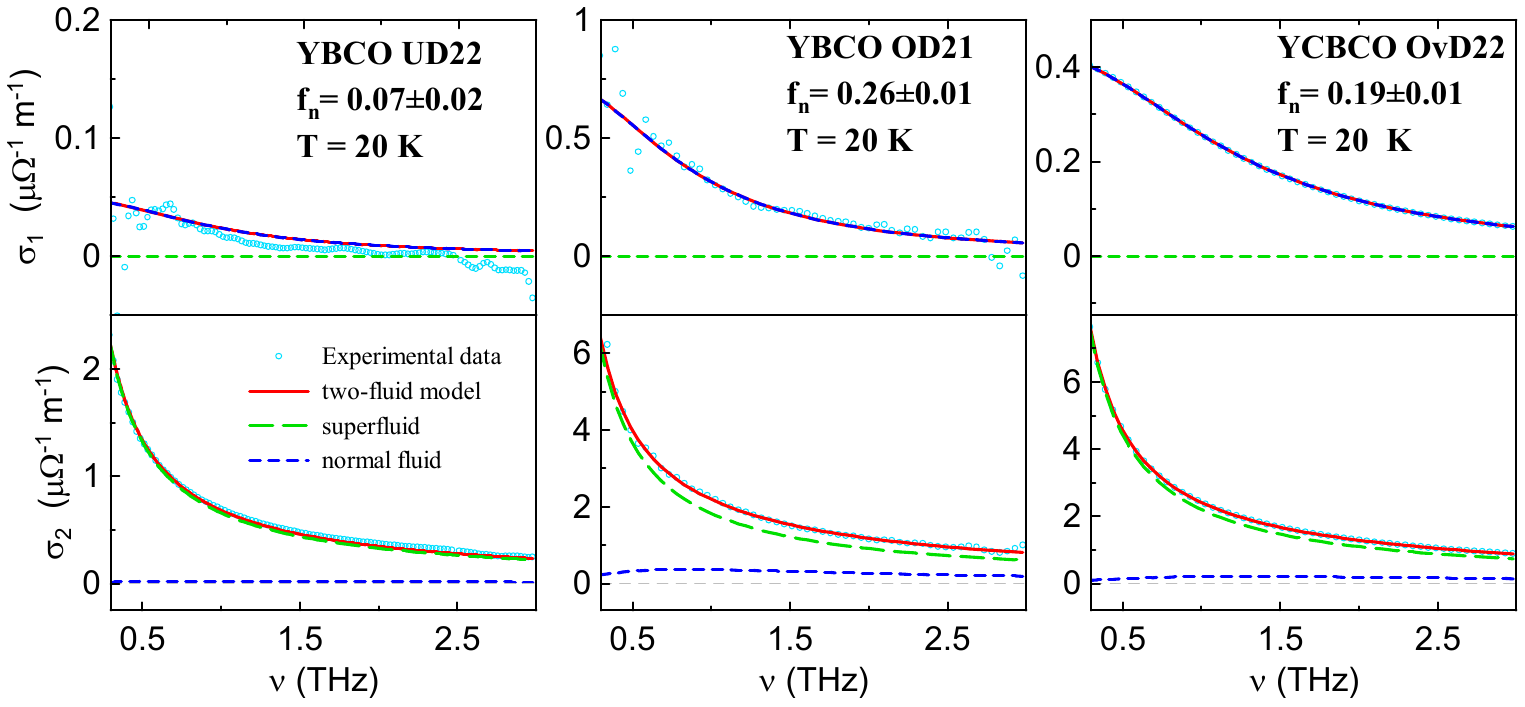}

 \caption{
The two-fluid model analysis of the low temperature conductivity for the three samples. The normal fluid and the superfluid contributions are shown by dashed lines.
 }
 \label{fig:2f}
 \end{figure}

In Fig~\ref{fig:2f}, we present a detailed two-fluid model analysis of the low temperature THz conductivity for the underdoped, optimally doped and overdoped samples. The total response from the two-fluid model (red solid line) has two components, one from the normal fluid (green dashed lines) and one from the superfluid (blue dashed line). The figure justifies the incomplete condensation for optimally doped and overdoped samples.